\documentclass[journal]{IEEEtran}
\usepackage[]{times}
\usepackage{epsfig}
\usepackage{subfigure}
\usepackage{bm,amsmath}
\usepackage{amssymb} 
\usepackage{graphicx}
\usepackage{subfigure}
\usepackage{multirow}
\usepackage{color}
\usepackage{url}
\usepackage{flushend}
\newcommand{\supp}{\mbox{supp}}
\newcommand{\cost}{\mbox{cost}}
\newtheorem{defn} {Definition}

\begin{document}
\title{Instanton-based Techniques for Analysis and Reduction of Error Floors of LDPC Codes}
\author{Shashi~Kiran~Chilappagari,~\IEEEmembership{Member,~IEEE,}
\thanks{Manuscript received October 1, 2008. Revised January 15, 2009 and March 6, 2009. S.~K.~Chilappagari [shashic@ece.arizona.edu] is with the Electrical and Computer Engineering Department, University of Arizona, Tucson, AZ, 85721, USA. }
~Michael Chertkov,~\IEEEmembership{Member,~IEEE,}
\thanks{M.~Chertkov [chertkov@lanl.gov] is with Theory Division \& CNLS, LANL, Los Alamos, NM, 87545 USA.}
~Mikhail G. Stepanov,
\thanks{M.~G.~Stepanov [stepanov@math.arizona.edu] is with the Department of Mathematics, University of Arizona, Tucson, AZ, 85721, USA.}
and ~Bane Vasic, ~\IEEEmembership{Senior Member,~IEEE}
\thanks{B.~Vasic [vasic@ece.arizona.edu] is with the Electrical and Computer Engineering Department, University of Arizona, Tucson, AZ, 85721, USA.}}

\maketitle
\thispagestyle{empty}
\begin{abstract}
We describe a family of instanton-based optimization methods developed recently for the analysis of the error floors of low-density parity-check (LDPC) codes. Instantons are the most probable configurations of the channel noise which result in decoding failures. We show that the general idea and the respective optimization technique are applicable broadly to a variety of channels, discrete or continuous, and variety of sub-optimal decoders. Specifically, we consider: iterative belief propagation (BP) decoders, Gallager type decoders, and linear programming (LP) decoders performing over the  additive white Gaussian noise channel (AWGNC) and the binary symmetric channel (BSC). 

The instanton analysis suggests that the underlying topological structures of the most probable instanton of the same code but different channels and decoders are related to each other. Armed with this understanding of the graphical structure of the instanton and its relation to the decoding failures, we suggest a method to construct codes whose Tanner graphs are free of these structures, and thus have less significant error floors.
\end{abstract}

\begin{keywords}
Low-density parity-check codes, Error Floor, Iterative Decoding, Linear Programming Decoding, Instantons, Pseudo-Codewords, Trapping Sets
\end{keywords}

\section{Introduction}\label{section1}
LDPC codes \cite{63Gallager}, \cite{99Mackay}, have been the focus of intense research over the past decade because they can approach theoretical limits of reliable transmission over various channels even when decoded by sub-optimal low complexity algorithms.

Two important classes of such algorithms are (i) iterative decoding algorithms, which include message passing algorithms (variants of the BP algorithm \cite{88Pearl} and Gallager type algorithms \cite{63Gallager}), and bit flipping algorithms \cite{76ZP,96SS} (serial and parallel), as well as (ii) the LP decoding algorithm \cite{05FWK}. Characterization of the error performance of sub-optimal algorithms (or simply decoders) is still an open problem, and has been addressed for both LDPC code ensembles, as well as for individual codes \cite{08RU}. Error performance of LDPC codes in the asymptotic limit of the code length is well characterized for a large class of sub-optimal decoders over different channels (the interested reader is referred to \cite{63Gallager,01RU,01RSU,04BRU} for general theory of message passing algorithms, \cite{76ZP,96SS,01BM,08Burshtein} for analysis of bit flipping algorithms and expander based arguments and \cite{05FS,07FMSSW,08DDKW} for analysis of the LP decoder).

A common feature of all the analysis methods used in deriving the asymptotic results is that the underlying assumptions hold in the limit of infinitely long code and/or are applicable to an ensemble of codes. Hence, they are of limited use for the analysis of a given finite length code. The performance of a code under a particular decoding algorithm is characterized by the bit-error-rate (BER) or the frame-error-rate (FER) curve plotted as a function of the signal-to-noise ratio (SNR). A typical BER/FER vs SNR curve consists of two distinct regions. At small SNR, the error probability decreases rapidly with SNR, with the curve looking like a \textit{waterfall}. The decrease slows down at moderate values turning into the \textit{error floor} asymptotic at very large SNR \cite{03Richardson}. This transient behavior and the error floor asymptotic originate from the sub-optimality of decoder, i.e., the ideal maximum-likelihood (ML) curve would not show such a dramatic change in the BER/FER with the SNR increase. While the slope of the BER/FER curve in the waterfall region is the same for almost all the codes in the ensemble, there can be a huge variation in the slopes for different codes in the error floor region \cite{08RU}. Since for sufficiently long codes the error floor phenomenon manifests itself in the domain unreachable by brute force Monte-Carlo (MC) simulations, analytical methods are necessary to characterize the FER performance.

Finite length analysis of LDPC codes is well understood for decoding over the binary erasure channel (BEC). The decoder failures in the error floor domain are governed by combinatorial structures known as stopping sets \cite{02DPRTU}. Stopping set distributions of various LDPC ensembles have been studied by Orlitsky \textit{et al.} (see \cite{05OVZ} and references therein for related works). Unfortunately, such a level of understanding of the decoding failures has not been achieved for other important channels such as the BSC and the AWGNC.

In this paper, we focus on the decoding failures of LDPC codes for iterative as well as LP decoders over the BSC and the AWGNC. Failures of iterative decoders for graph based codes were first studied by Wiberg \cite{96Wiberg} who introduced the notions of \textit{computation trees} and \textit{pseudo-codewords}. Subsequent analysis of the computation trees was carried out by Frey \textit{et al.} \cite{01FKV} and Forney \textit{et al.}\cite{01Forney}. The failures of the LP decoder can be understood in terms of the vertices of the so-called \textit{fundamental polytope} which are also known as pseudo-codewords \cite{05FWK}. Vontobel and Koetter \cite{05VK} introduced a theoretical tool known as graph cover approach and used it to establish connections between the LP and the message passing decoders using the notion of the fundamental polytope. They showed that the pseudo-codewords arising from the Tanner graph covers are identical to the pseudo-codewords of the LP decoder. Vontobel and Koetter \cite{04VK} also studied the relation between the LP  and the min-sum decoders.

For iterative decoding on the AWGNC, MacKay and Postol \cite{03MP} were the first to discover that certain ``near codewords'' are to be blamed for  the high error floor in the Margulis code. Richardson \cite{03Richardson} reproduced their results and developed a computation technique to predict the  performance of a given LDPC code in the error floor domain. He characterized the troublesome noise configurations leading to the error floor using combinatorial objects termed trapping sets and described a technique (of a Monte-Carlo importance sampling type) to evaluate the error rate associated with a particular class of trapping sets. The method from \cite{03Richardson} was further refined for the AWGNC by Stepanov \textit{et al.} \cite{05SCCV} who introduced the notion of \textit{instantons}. In a  nutshell, an instanton is a configuration of the noise which is positioned in between a codeword (say zero codeword) and another pseudo-codeword (which is not necessarily a codeword). Incremental shift (allowed by the channel) from this configuration toward the zero codeword leads to correct decoding (into the zero-codeword) while incremental shift in an opposite direction leads to a failure.  In principle, one can find this dangerous configuration of the noise by exploring the domain of correct decoding surrounding the zero codeword, and finding borders of this domain -- the so-called error-surface. If the channel is continuous,  the error-surface consists of continuous patches while configuration of the noise maximizing the error probability over a patch is called an instanton. The term instanton introduced initially in the context of disordered systems is also known under the names of \textit{saddle-point} or \textit{optimal fluctuation}, and is common in modern theoretical physics  (see \cite{05SCCV} and references therein).

As stated above, the instantons  that affect the decoder performance in the error floor region are extremely rare, and hence identifying and enumerating them is a challenging task. However, once this difficulty is overcome, the knowledge of the trapping set/pseudo-codeword distribution can be used to evaluate the performance of the code. It can also be used to guide optimization of the code and design of improved decoding strategies.  In this paper, we focus on the methods used to identify the most relevant noise configurations for various decoders and channel models.

Previous investigation of the problem include the work by Kelley and Sridhara \cite{07KS} who studied pseudo-codewords arising from graph covers and derived bounds on the minimum pseudo-codeword weight in terms of the girth and the minimum left-degree of the underlying Tanner graph. The bounds were further investigated by Xia and Fu \cite{08XF}. Smarandache and  Vontobel \cite{07SV} found pseudo-codeword distributions for the special cases of codes from Euclidean and projective planes. Pseudo-codeword analysis has also been extended to the convolutional LDPC codes by Smarandache \textit{et al.} \cite{06SPVC}. Milenkovic \textit{et al.} \cite{07MSW} studied the asymptotic distribution of trapping sets in regular and irregular ensembles. Wang \textit{et al.} \cite{06WKP} proposed an algorithm to exhaustively enumerate certain trapping sets.

Chernyak \textit{et al.} \cite{04CCSV} and Stepanov \textit{et al.} \cite{05SCCV} suggested to pose this problem of finding the instantons as a special optimization problem. This optimization method was built in the spirit of the general methodology, borrowed from statistical physics, guiding exploration of rare events which contribute the most to the BER/FER. The optimization method allowed to discover in \cite{05SCCV}, the set of most probable instantons for the AWGNC and iterative decoder. The operational utility of the method was illustrated on some number of moderate size examples and  dependence of the instanton structure on the number of iterations was observed. The general optimization method was substantially improved and refined in \cite{08CS} for the LP decoder over continuous channels (with main enabling example chosen to be the AWGNC). The pseudo-codeword-search (PCS) algorithm of \cite{08CS} was essentially exploring in an iterative way the Wiberg formula, treating an instanton configuration as a median between a pseudo-codeword and the zero-codeword.

It was shown empirically that, initiated with a sufficiently noisy configuration, the algorithm converges to an instanton in sufficiently small number of steps, independent or weakly dependent on the code size. Repeated multiple times, the method outputs the set of instanton configurations which can further be used to estimate the BER/FER performance in the transient and error floor domain.  The definition of the instantons and the instanton search method were extended in \cite{08CCV} to the BSC. In this special case, the instanton search algorithm is provably efficient, in the sense that it outputs an instanton in small number of steps, and that the weight of the pseudo-codeword found in the intermediate steps is monotonically decreasing. (See also \cite{07Vontobel} for an exhaustive list of references for this and related subjects.)

In this paper, we discuss failures of iterative decoders (specifically the BP algorithm and the Gallager A/B algorithms) as well as LP decoding over the BSC and the AWGNC. We explain the notion of instanton and elaborate on the connections between instantons and trapping sets as well as pseudo-codewords. We then describe algorithms to search for instantons. By using the $[155,64,20]$ Tanner code \cite{01TSF} as an enabling example, we illustrate the performance of the instanton search technique outputting the set of  most probable instantons. By identifying that all decoding failures can be attributed to the presence of certain subgraphs, we construct a code avoiding this subgraph and show that this code outperforms the original code. Throughout the paper, we focus on the BSC and the AWGNC and while the underlying approach is similar for both channels, rigorous statements can be made for the BSC \cite{08CCV}, while the respective AWGNC statements come from experiments only.

The rest of the paper is organized as follows. In Section \ref{section2}, we introduce the notation and provide the required  background material. The notions of decoding failures and instantons are discussed in Section \ref{section3}, followed by a description of instanton search algorithms for different decoders in Section \ref{section4}. We illustrate numerical results in Section \ref{section5} and conclude in Section \ref{section6}.
\section{Preliminaries}\label{section2}
\subsection{LDPC Codes}
LDPC codes belong to the class of linear block codes which can be defined by sparse bipartite graphs \cite{81Tanner}. The Tanner graph \cite{81Tanner} $G$ of an LDPC code $\mathcal{C}$  is a bipartite graph with two sets of nodes: the set of variable nodes $V=\{1,2,\ldots,n\}$ and the set of check nodes $C=\{1,2,\ldots,m\}$. The check nodes (variable nodes resp.) connected to a variable node (check node resp.) are referred to as its neighbors. The set of neighbors of a node $u$ is denoted by $\mathcal{N}(u)$. The degree $d_u$ of a node $u$ is the number of its neighbors. A vector $\mathbf{v}=(v_1,v_2,\ldots,v_n)$ is a codeword if and only if for each check node, the modulo two sum  of its neighbors is zero. An $(n,d_v,d_c)$ regular LDPC code has a Tanner graph with $n$ variable nodes each of degree $d_v$ and $nd_v/d_c$ check nodes each of degree $d_c$. This code has length $n$ and rate  $r \geq 1-d_v/d_c$ \cite{63Gallager}. It should be noted that the Tanner graph is not uniquely defined by the code and when we say the Tanner graph of an LDPC code, we only mean one possible graphical representation.

\subsection{Channel Assumptions}

We assume that a binary codeword $\mathbf{y}$ is transmitted over a noisy channel and is received as $\mathbf{\hat{y}}$. The support of a vector $\mathbf{y}=(y_1,y_2,\ldots,y_n)$, denoted by $\supp(\mathbf{y})$, is defined as the set of all positions $i$ such that $y_i\neq 0$. In this paper, we consider binary input memoryless channels with discrete or continuous output alphabet. As the channel is memoryless, we have
\[
\Pr(\hat{\mathbf{y}}| \mathbf{y}) = \prod_{i \in V} \Pr(\hat{y}_i| y_i)
\]
and hence can be characterized by $\Pr(\hat{y}_i|y_i)$, the probability that $\hat{y}_i$ is received given that $y_i$ was sent.
The negative log-likelihood ratio (LLR) corresponding to the variable node $i \in V$ is given by
\[
\gamma_i=\log\left(\frac{\Pr(\hat{y}_i| y_i=0)}{\Pr(\hat{y}_i| y_i=1)}\right).
\]
Two binary input memoryless channels of interest are the BSC with output alphabet $\{0,1\}$ and the AWGNC with output alphabet $\mathbb{R}$. On the BSC with transition probability $\epsilon$,  every transmitted bit $y_i \in \{0,1\}$ is flipped \footnote{The event of a bit changing from $0$ to $1$ and vice-versa is known as flipping.} with probability $\epsilon$ and is received as $\hat{y}_i \in \{0,1\}$. Hence, we have
\[
\gamma_i = \left \{ \begin{array}{cl} \log \left( \frac{1-\epsilon}{\epsilon}\right) & \mbox{~if~} \hat{y}_i=0 \\
\log \left( \frac{\epsilon}{1-\epsilon}\right) & \mbox{~if~} \hat{y}_i=1 \end{array}\right.
\]

For the AWGNC, we assume that each bit $y_i \in \{0,1\}$ is modulated using binary phase shift keying (BPSK) and transmitted as $\overline{y}_i = 1-2y_i$ and is received as $\hat{y}_i = \overline{y}_i + n_i$, where $\{n_i\}$ are i.i.d. $N(0,\sigma^2)$ random variables. Hence, we have
\[
\gamma_i = \frac{2\hat{y}_i}{\sigma^2}.
\]

\subsection{Decoding Algorithms}
\subsubsection{Message Passing Decoders}
Message passing decoders operate by passing messages along the edges of the Tanner graph representation of the code. Gallager in \cite{63Gallager} proposed two simple binary message passing algorithms for decoding over the BSC; Gallager A and Gallager B. There exist a large number of message passing algorithms (the BP algorithm, the min-sum algorithm, quantized decoding algorithms, decoders with erasures \cite{01RU} to name a few )  in which the messages belong to a larger alphabet.

Let $\mathbf{\hat{y}}=(\hat{y}_1,\hat{y}_2,\ldots,\hat{y}_n)$, an $n$-tuple be the input to the decoder. Let $\omega^{(k)}_{i \to \alpha}$ denote the message passed by a variable node $i \in V$ to its neighboring check node $\alpha \in C$ in the $k^{th}$ iteration and $\varpi^{(k)}_{\alpha \to i}$ denote the message passed by a check node $\alpha$ to its neighboring variable node $i$. Additionally, let $\varpi^{(k)}_{* \to i}$ denote the set of all incoming messages to variable node $i$ and $\varpi^{(k)}_{* \backslash \alpha \to i}$ denote the set of all incoming messages to variable node $i$ except from check node $\alpha$. 

\noindent \textbf{Gallager A/B Algorithm:}
The Gallager A/B algorithms are hard-decision-decoding algorithms in which all the messages are binary. With a slight abuse of the notation, let $|\varpi_{* \to i}=m|$ denote the number of incoming messages to $i$ which are equal to $m \in \{0,1\}$. Associated with every decoding round $k$ and variable degree $d_i$ is a threshold $b_{k,d_i}$. The Gallager B algorithm is defined as follows.
\begin{eqnarray}
\omega_{i \to \alpha}^{(0)}&=&\hat{y}_i \nonumber \\
\varpi^{(k)}_{\alpha \to i}&=& \left(\sum_{j \in \mathcal{N}(\alpha)\backslash i} \omega^{(k-1)}_{j \to \alpha}\right) \mbox{mod } 2 \nonumber \\
\omega^{(k)}_{i \to \alpha} &=& \left \{ \begin{array}{cl}
												1, & \mbox{if } |\varpi^{(k)}_{* \backslash \alpha \to i}=1| \geq b_{k,d_i}\\
											0, & \mbox{if } |\varpi^{(k)}_{* \backslash \alpha \to i}=0| \geq b_{k,d_i}\\
											\hat{y}_i, & \mbox{otherwise}  \end{array}\right. \nonumber 
\end{eqnarray}
The Gallager A algorithm is a special case of the Gallager B algorithm with $b_{k,d_i}=d_i-1$ for all $k$. At the end of each iteration, a decision on the value of each variable node is made based on all the incoming messages and possibly the received value.

\noindent \textbf{The BP Algorithm:}
A soft-decision-decoding algorithm, which is the best possible one if the messages are calculated locally in the Tanner graph of the code, is the BP algorithm (also known as the sum-product algorithm). With a moderate abuse of notation, the messages passed in the BP algorithm are described below:
\begin{eqnarray}
 \omega^{(0)}_{i \to \alpha} & = & \gamma_i \nonumber \\
 \varpi^{(k)}_{\alpha \to i} & = & 2\tanh^{-1} \left( \prod_{j \in \mathcal{N}(\alpha)\backslash i} \tanh \left(
   \frac{1}{2}\omega^{(k-1)}_{j \to \alpha} \right) \right) \nonumber \\
 \omega^{(k)}_{i \to \alpha} & = & \gamma_i + \sum_{\delta \in \mathcal{N}(i)\backslash \alpha} \varpi^{(k)}_{\delta \to i} \nonumber
\end{eqnarray}
The result of decoding after $k$ iterations, denoted by $\mathbf{x}^{(k)}$, is determined by the sign of $m_i^{(k)} = \gamma_i + \sum_{\alpha \in \mathcal{N}(i)} \varpi^{(k)}_{\alpha \to i}$. If $m_i^{(k)} > 0$ then $x_i^{(k)}=0$, otherwise $x_i^{(k)}=1$.

In the limit of high SNR, when the absolute value of the messages is
large, the BP algorithm becomes the min-sum algorithm, where the message
from the check $\alpha$ to the bit $i$ looks like:
\begin{eqnarray}
 \varpi^{(k)}_{\alpha \to i} & = & \min \big| \omega^{(k-1)}_{*
   \backslash i \to \alpha} \big| \cdot \prod_{j \in \mathcal{N}(\alpha)\backslash i} {\rm sign} \big(
   \omega^{(k-1)}_{j \to \alpha} \big) \nonumber
\end{eqnarray}

The min-sum algorithm has a property that the Gallager A/B and the LP
decoders also possess --- if we multiply all the likelihoods $\gamma_i$ by a
factor, all the decoding would proceed as before and would produce the
same result. Note that we do not have this ``scaling'' in the BP
algorithm.

To decode the message in complicated cases (when the message distortion
is large) we may need a large number of iterations, although typically a
few iterations would be sufficient. To speed up the decoding process one
may check after each iteration whether the output of the decoder is a
valid codeword, and if so halt decoding.

\subsubsection{Linear Programming Decoder}
The ML decoding of the code $\mathcal{C}$ allows a convenient LP formulation in
terms of the \textit{codeword polytope} $\mbox{poly}(\mathcal{C})$ \cite{05FWK} whose
vertices correspond to the codewords in $\mathcal{C}$. The ML-LP decoder finds
$\mathbf{f}=(f_1,\ldots,f_n)$ minimizing the cost function
$\sum_{i=1}^{n}\gamma_if_i$ subject to the $\mathbf{f}\in
\mbox{poly}(\mathcal{C})$ constraint. The formulation is compact but
impractical as the number of constraints is exponential in the code
length.

Hence a \textit{relaxed} polytope is defined as the intersection of all the
polytopes associated with the local codes introduced for all the checks of the
original code. Associating $(f_1,\ldots,f_n)$ with bits of the code we require
\begin{equation}\label{eq1}
0 \leq f_i \leq 1, ~~\forall i \in V
\end{equation}
For every check node $\alpha$, let $\mathcal{N}(\alpha)$ denote the set of variable nodes which are
neighbors of $\alpha$. Let $E_\alpha=\{T \subseteq \mathcal{N}(\alpha): |T| \mbox{~is even}\}$. The
polytope $Q_\alpha$ associated with the check node $\alpha$ is defined as the set of
points $(\mathbf{f},\mathbf{w})$ for which the following constraints hold
\begin{eqnarray}
&0 \leq w_{\alpha,T} \leq 1,& \forall T \in E_\alpha \\
&\sum_{T \in E_\alpha} w_{\alpha,T}=1& \\
&f_i=\sum_{T \in E_\alpha, T \ni i }w_{\alpha,T},& \forall i \in \mathcal{N}(\alpha) \label{eq4}
\end{eqnarray}
Now, let $Q=\cap_\alpha Q_\alpha$ be the set of points $(\mathbf{f},\mathbf{w})$ such that
(\ref{eq1})-(\ref{eq4}) hold for all $\alpha \in C$. (Note that $Q$, which is also referred to as the fundamental polytope \cite{03KV,05VK}, is a function of the Tanner graph $G$ and consequently the parity-check matrix $H$ representing the code $\mathcal{C}$.) The linear code linear program (LCLP) can be stated as
\[
\min\limits_{(\mathbf{f},\mathbf{w})} \sum_{i \in V}\gamma_i f_i, \mbox{~s.t.~} (\mathbf{f},\mathbf{w}) \in Q.
\]
For the sake of brevity, the decoder based on the LCLP is referred to in the
following as the LP decoder. A solution $(\mathbf{f},\mathbf{w})$ to the
LCLP such that all $f_i$s and $w_{\alpha,T}$s are integers is known as an integer
solution. The integer solution represents a codeword \cite{05FWK}. It was also
shown in \cite{05FWK} that the LP decoder has the ML certificate, i.e., if the
output of the decoder is a codeword, then the ML decoder would decode into the
same codeword. The LCLP can fail, generating an output which is not a codeword.

It is appropriate to mention here that the LCLP can be viewed as the zero
temperature version of  BP-decoder looking for the global minimum of the
so-called Bethe free energy functional \cite{03WJ}.

\section{Decoding Failures and Instantons}\label{section3}

To characterize the performance of a coding/decoding scheme for linear codes over any output symmetric channel, one can assume, without loss of generality, the transmission of the all-zero-codeword, i.e. ${\bm y}={\bm 0}$, when the decoding algorithm satisfies certain symmetry conditions (see Definition 1 and Lemma 1 in \cite{01RU}). The iterative decoding algorithms that we consider in this paper satisfy these symmetry conditions. The assumption of the transmission of the all-zero-codeword also holds for the LP decoding of linear codes on output symmetric channels, as the polytope $Q$ is highly symmetric and looks exactly the same from any codeword (see \cite{05FWK} for proof). Henceforth, we assume that ${\bm y}={\bm 0}$.

A decoding failure is said to have occurred if the output of the decoder is not equal to the transmitted codeword (all-zero-codeword). Probability of a decoder failure, or the frame error rate as a function of the SNR $s (=E_b/N_0)$ can be expressed as:
\begin{eqnarray}
FER(s)=\sum_{\hat{\bm y}} P_{s}(\hat{\bm y})\theta(\hat{\bm y}),
\label{FER}
\end{eqnarray}
where the sum goes over all the possible outputs of the channel for the zero-codeword input. In case of a continuous output channel, the sum becomes an integral: $\sum\to\int d\hat{\bm y}$, and the channel probability mass function becomes a probability density function: $\int d\hat{\bm y} P_{s}(\hat{\bm y})=1$.
$\theta(\hat{\bm y})$ in Eq.~(\ref{FER}) is defined to be zero, in the case of successful decoding, and is unity in the case of failure. $P_{s}(\hat{\bm y})$ is the probability of observing $\hat{\bm y}$ at the output of a channel characterized by the SNR $s$ \footnote{Note that for the BSC, the transition probability $\epsilon$ is a measure of the SNR. For code rate $r$ and BPSK modulated transmission over the AWGNC with noise variance $\sigma^2$, we have $E_b/N_0=1/(2r\sigma^2)$.}.

Calculating the above sum/integral exactly is not feasible, and the instanton-based approach consists of approximating the sum/integral by a finite number of terms corresponding to the most probable failures -- the instantons. This approximation becomes asymptotically exact in the limit of large SNR, while at smaller SNRs, more terms are needed to obtain accurate approximation for the FER. Note that the details of the approximate evaluations are different for discrete and continuous channels. In the discrete case, the number of terms is finite. We account for the $k$-most probable configurations, and $FER(s)\approx \sum_{\beta=1}^k {\cal N}_\beta P_{s}(\hat{\bm y}_\beta)$, where the multiplicity factor ${\cal N}_\beta$ counts the number of instantons equivalent under bit permutations. For continuous channels, an instanton is a stationary point of the respective integrand. By stationary point, we mean the local maximum of the noise probability density function. Hence for the AWGNC, instanton is defined as the noise configuration with minimal (probably locally) value of the $L^2$ norm of $\hat{\mathbf{y}}$ that leads to a decoding failure. The $L^2$ norm of a vector $\hat{\mathbf{y}}$ is equal to $\sqrt{\sum_{i \in V}\hat{y_i}^2}$. Note that for the AWGNC, smaller the $L^2$ norm, the more probable the noise configuration is.

\begin{figure}
\centering
\includegraphics[width=2.2in]{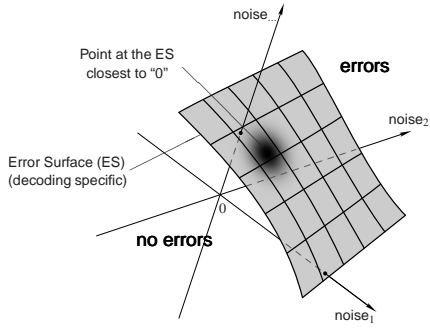}
\caption{Illustration of error surface.}
\label{Errorsurface}
\end{figure}

The FER approximation should also include, in addition to the multiplicities, the curvature corrections around the stationary point (e.g. within Gaussian approximation) \cite{04CCSV,06SC}. In other words,
$ FER(s)\approx \sum_{\beta=1}^k {\cal N}_\beta {\cal C}_s(\hat{\bm y}_\beta) P_s(\hat{\bm y}_\beta)$,  where ${\cal C}_s(\hat{\bm y}_\beta)$ is the curvature factor. The multiplicities of the instantons are determined by the symmetry group of the code, and if nothing is known about it, one may assume that the group is trivial and all multiplicities are equal to $1$. In continuous case, the curvatures are determined by the geometry of the error surface in the vicinity of the instanton (this subject was not studied numerically so far, as the most important information about the instanton is its weight which determines the slope of FER {\it vs.} SNR curve in the asymptotic). Intuitively, in the case of the AWGNC and $s\to\infty$, ${\cal C}_s(\hat{\bm y}_\beta)=O(1/\sqrt{s})$, the decay of the noise correlations is exponential along one direction (orthogonal to the error surface) and quadratic along the remaining $N-1$ components of the noise vector (see Fig. \ref{Errorsurface} for an illustration of the error surface). 

Consistent with the above statements, instantons $\hat{\bm y}_s$ can be also defined as special configurations of the noise resulting in decoding failures such that any incremental (and channel specific) shift of the noise toward the zero-codeword results in correct decoding. It is thus useful to also introduce a respective output, $\tilde{\bm y}_s=\mbox{dec}(\hat{\bm y}_s)$, called a pseudo-codeword.  It should be noted that this informal definition of the pseudo-codewords is generic  and applicable to any channel and decoder. While the output for the LP decoder is well defined and does not suffer from  numerical issues, the iterative decoder can exhibit oscillations i.e., the bits which are decoded wrongly can differ from one iteration to another. As a way to streamline the description of decoding failures in the presence of rounding and iterative uncertainties, Richardson \cite{03Richardson} suggested a proxy notion of the trapping set, which is a combinatorial object that accounts for the decoder output over iterations. In the subsequent discussion, we formally define trapping sets and pseudo-codewords and also provide some BSC-specific definitions. If an instanton of a channel/decoder is known, the respective pseudo-codeword can be easily found,  and conversely if a pseudo-codeword is given (i.e. we know for sure that there exists a configuration of the noise which is sandwiched in between the pseudo-codeword and the zero-codeword) the respective instanton can be restored. In fact,  this inversion is in the core of the pseudo-codeword/instanton search algorithms discussed in Section \ref{section4}. 

\subsection{Trapping Sets for Iterative Decoders} 
In practice, we assume that the iterative decoder performs a finite  number $D$ of iterations. Let $\mathbf{\hat{y}}=(\hat{y}_1,\hat{y}_2,\ldots,\hat{y}_n)$ be a vector which is the input to the iterative decoder and let $\mathbf{x}^{(k)}=(x_1^{(k)},x_2^{(k)},\ldots, x_n^{(k)})$, $k \leq D$ be the output binary vector at the $k^{th}$ iteration. A variable node $i$ is said to be \textit{eventually correct} if there exists a positive integer $K$ such that for all $k \geq K$, $x^{(k)}_i =0$ \cite{03Richardson}. Formally, a decoder failure is said to have occurred if there does not exist $k$ such that $\supp(\mathbf{x}^{(k)})=\emptyset$ \cite{03Richardson}.
\begin{defn}\cite{03Richardson}(Trapping sets for iterative decoders:)
For an input $\mathbf{\hat{y}}$, let $\mathbf{T}(\mathbf{\hat{y}})$ denote	the set of variable nodes that are not eventually correct. If $\mathbf{T}(\mathbf{\hat{y}})\neq \emptyset$, then $\mathbf{T}(\mathbf{\hat{y}})$ is a trapping set. If $a=|\mathbf{T}(\mathbf{\hat{y}})|$ and $b$ is the number of odd degree check nodes in the sub-graph induced by $\mathbf{T}(\mathbf{\hat{y}})$, we say $\mathbf{T}(\mathbf{\hat{y}})$ is an $(a,b)$ trapping set.
\end{defn}

For the BSC, since the input to the decoder as well as the messages passed are discrete, it is easier to define instantons in terms of number of bits flipped in the input to the decoder. The instantons with least number of flips will be the most dominant in the error floor region. We formalize this intuition below.

\begin{defn}(Critical number for Gallager A/B algorithm)
Let $\mathcal{T}$ be a trapping set for the Gallager A/B algorithm and let $\mathbf{\hat{y}} \in GF(2)^n$. Let $\mathbf{Y}(\mathcal{T})=\{\mathbf{\hat{y}}|\mathbf{T}(\mathbf{\hat{y}})=\mathcal{T}\}$. The critical number $m(\mathcal{T})$ of trapping set $\mathcal{T}$ for the Gallager A/B algorithm is the minimum number of variable nodes that have to be initially in error for the decoder to end up in the trapping set $\mathcal{T}$, i.e.,
\[
\displaystyle m(\mathcal{T})=\min_{\mathbf{Y}(\mathcal{T})}{|\supp(\mathbf{\hat{y}})|}.
\]
\end{defn}

The most relevant trapping set in the error floor region is the trapping set with the least critical number. 
\begin{defn}(Instanton for Gallager A/B over the BSC)
An instanton is a binary vector $\bm{i}$ such that $\mathbf{T}(\bm{i}) = \mathcal{T}$ for some trapping set $\mathcal{T}$ and for any binary vector $\bm{r}$ such that $\supp(\bm{r}) \subset \supp(\bm{i})$, $ \mathbf{T}(\bm{r})=\emptyset$. The size of an instanton is the cardinality of its support.
\end{defn}

Given a trapping set, one can  consider vectors whose support is a subset of the trapping set as input to the decoder and see if such vectors are instantons. While rigorous statements cannot be made about finding smallest size instantons, the above method gives instantons in most of the cases (see \cite{06CSV} for some illustrations). Intuitively, this seems reasonable as we do not expect inputs to the decoder which do not have errors in variable nodes involved in a trapping set to end up in a trapping set. 

\subsection{Pseudo-codewords for LP Decoders} 
In contrast to the iterative decoders, the output of the LP decoder is well defined in terms of pseudo-codewords.

\begin{defn}\cite{05FWK}
An \textit{integer pseudo-codeword} is a vector $\mathbf{p}=(p_1,\ldots,p_n)$ of
non-negative integers such that, for every parity check $\alpha \in C$,
the neighborhood $\{p_i: i \in \mathcal{N}(\alpha)\}$ is a sum of local codewords.
\end{defn}
The interested reader is referred to Section V in \cite{05FWK} for more details and examples. Alternatively, one may choose to define a \textit{re-scaled pseudo-codeword},
$\mathbf{p}=(p_1,\ldots,p_n)$ where $0 \leq p_i \leq 1, \forall i
\in V$, simply equal to the output of the LCLP. In the following, we
adopt the re-scaled definition. The cost associated with LP decoding of a vector $\mathbf{\hat {y}}$ to a pseudo-codeword $\mathbf{p}$ is given by
\begin{equation}
\cost(\mathbf{\hat{y}},\mathbf{p})= \sum_{i \in V} \gamma_i p_i. \nonumber
\end{equation}
For an input $\mathbf{\hat{y}}$, the LP decoder outputs the pseudo-codeword $\mathbf{p}$ with minimum $\cost(\mathbf{\hat{y}},\mathbf{p})$. Since the cost associated with LP decoding of $\mathbf{\hat {y}}$ to the all-zero-codeword is zero, a decoder failure occurs on the input $\mathbf{\hat{y}}$ if and only if there exists a pseudo-codeword $\mathbf{p}$ with $\cost(\mathbf{\hat{y}},\mathbf{p}) \leq 0$.

A given code $\mathcal{C}$ may have different Tanner graph representations and consequently potentially different fundamental polytopes. Hence, we refer to the pseudo-codewords as corresponding to a particular Tanner graph $G$ of $\mathcal{C}$. 

\begin{defn}\cite{01Forney}
Let $\mathbf{p}=(p_1,\ldots,p_n)$ be a pseudo-codeword distinct from the all-zero-codeword of the code $\mathcal{C}$ represented by Tanner graph $G$ .  Then, the \textit{pseudo-codeword weight} of $\mathbf{p}$ is defined as follows:
\begin{itemize}
\item $w_{BSC}(\mathbf{p})$ for the BSC is
\[
w_{BSC}(\mathbf{p})=\left\{ \begin{array}{cl}
2e,& \mbox{~if~} \sum_{e}p_i=\left(\sum_{i \in V}p_i\right)/2; \\
2e-1,&\mbox{~if~} \sum_{e}p_i>\left(\sum_{i \in V}p_i\right)/2. \end{array}\right.
\]
where $e$ is the smallest number such that the sum of the $e$ largest $p_i$s is at least $\left(\sum_{i \in V}p_i\right)/2$.
\item $w_{AWGN}(\mathbf{p})$ for the AWGNC is
\[
w_{AWGN}(\mathbf{p})=\frac{(p_1+p_2+\ldots + p_n)^2}{(p_1^2+p_2^2+\ldots+ p_n^2)}
\]
\end{itemize}
\end{defn}
The minimum pseudo-codeword weight of $G$ denoted by $w_{min}^{BSC/AWGN}$ is the minimum over all the non-zero pseudo-codewords of $G$.

We now give definitions specific to the BSC. \begin{defn}(Median for LP decoding over the BSC) The median noise vector (or simply the median) $M(\mathbf{p})$ of a pseudo-codeword $\mathbf{p}$ distinct from the all-zero-codeword is a binary vector with support $S=\{i_1,i_2,\ldots,i_e\}$, such that $p_{i_1},\ldots,p_{i_e}$ are the $e(=\lceil \left(w_{BSC}(\mathbf{p})+1\right)/2\rceil)$ largest components of $\mathbf{p}$.
\end{defn}

Note that for input $\mathbf{\hat{y}}=M(\mathbf{p})$ for some non-zero pseudo-codeword $\mathbf{p}$, we have $\cost(\mathbf{\hat{y}},\mathbf{p})\leq 0$ and hence leads to a decoding failure (the output of the decoder, however, need not be the pseudo-codeword we start with).

\begin{defn}(Instanton for LP decoding over the BSC)
The BSC \textit{instanton} $\mathbf{i}$ is a binary vector with the following
properties: (1) There exists a pseudo-codeword $\mathbf{p}$ such that
$\cost(\mathbf{i},\mathbf{p})\leq \cost(\mathbf{i},\mathbf{0})=0$; (2) For any binary
vector $\mathbf{r}$ such that $\supp(\mathbf{r}) \subset \supp(\mathbf{i})$,
there exists no pseudo-codeword with $\cost(\mathbf{r},\mathbf{p})\leq 0$. The size of an instanton is the cardinality of its support.
\end{defn}

An attractive feature of LP decoding over the BSC is that any input whose support contains an instanton leads to a decoding failure (which is not the case for Gallager A decoding over the BSC) \cite{08CCV}. This important property is in fact used in searching for instantons.

To summarize, evaluating FER vs SNR approximately reduces to finding the set of most probable instantons and evaluating their probabilities, multiplicities and, in the continuous case, also respective curvatures. Specifically, for LP decoding over the BSC and the Gallager algorithm, the slope of the FER curve in the error floor region is equal to the cardinality of the smallest size instanton (see \cite{08ICV} for a formal description). Understanding that the knowledge of the instantons allows efficient approximation of FER vs SNR dependence (which is our main task), we now discuss approaches to finding the set of instantons for a given error-correction setting in Section \ref{section4}.

\section{Searching for Instantons}\label{section4}

As explained above in Section \ref{section3}, instantons that control the large SNR asymptotic of the FER are the most probable noise configurations corresponding to decoder failures. Stated this way the problem of finding an instanton becomes an optimization problem, and all the remaining details of this section are related to efficient implementation of this, generally difficult, optimization problem.

\subsection{Instanton search for iterative decoding over continuous channels}
A straightforward optimization method for finding instantons in the case of a continuous channel is based on the standard (amoeba)  optimization \cite{92PBTV} and was discussed by Stepanov and Chertkov in \cite{06SC}. The main idea of the direct technique is as follows. One draws randomly a unit length configuration of the noise and finds a scale-up value which positions the re-scaled configuration of the noise exactly at the error-surface. Thus, incremental increase/decrease of the rescaling factor leads to decoding failure or recovery. Such a configuration and its probability are recorded, and this operation is repeated $(N-2)$ times,  thus generating $N-1$ vertices of a simplex with respective probabilities assigned. Then, aiming to find a more probable point in the interior of the simplex, the current point is transformed according to the standard amoeba rules. The process is repeated until the size of the simplex becomes smaller than a preset accuracy, and the resulting most probable configuration outputs an instanton. The whole optimization is repeated multiple number of times, each time generating an instanton. The main advantage of the method is in its generality (it can be used for any continuous channel and any soft decoding algorithm). However, implementing this method is costly. Although one can use amoeba optimization method for LP decoding too, because of a certain property of LP decoding (it is easy to find an instanton (noise realization) corresponding to the output of the decoding which is a pseudo-codeword), the PCS method described in Section \ref{section4c} is a lot more effective.  

The instanton-amoeba method easily finds the instantons for a code if the number of iterations in decoding is not large (less than $20$). Increase of the number of iterations, $n_{\mbox{\scriptsize it}}$, simply means longer computations. The other more important effect is associated with enhancement of irregular, stochastic component in decoding observed with $n_{\mbox{\scriptsize it}}$ increase. One finds that already a slight variation in the noise can drastically change results. That makes the function that we have to optimize very irregular, which dramatically slows down the optimization process.

In the case of large number of iterations in the decoder, with the check for a codeword in each iteration, it is not easy to come up with good starting point for the amoeba. The configurations from amoeba with small number of iterations (when the method is quite effective) are not very useful as the decoder eventually outputs a codeword. The following two ways to find such configurations were developed. Both are based on observations from numerical experiments.

1) Input an instanton for LP decoder to the min-sum iterative decoder. The instantons for LP decoding (as they have low weight)  serve as good seeding noise configurations for amoeba, as they are in erroneous domain in noise space even for a decoder with a very large number of iterations.  

2) Limit the noise configuration on bits where the instanton for low number of iterations is supported. Work then with an optimization problem on these bits only, setting the noise value on all other bits to zero. In this way, the number of variables is much lower, so the optimization procedure is a lot easier to proceed with. The smaller is the dimension of the space in which the amoeba optimization is done, the easier is the problem. The instantons for low number of iterations usually have noise in a few bit locations. One can hope that if one increases the noise level on these selected bits (while keeping the noise at all other bits being exactly zero) then the noise configuration will ``survive'' a lot more iterations. The $12.45$ weight instanton (supported by $12$ bits) for AWGNC and $410$ iterations decoder that is reported in \cite{06SC} was found this way.

\subsection{Instanton search for Gallager A/B decoders over the BSC}
In contrast to iterative decoding with continuous alphabet, the trapping sets and instantons for the Gallager A/B decoder can be found using certain combinatorial considerations which were first pointed out by Richardson \cite{03Richardson} and later investigated in detail in \cite{06CSV,08CNVM1,08CNVM2,08CV}. The trapping sets for Gallager type decoders are closely related to trapping sets for the bit flipping decoders.

\subsection{Instanton search for LP decoding over the AWGNC}\label{section4c}
For the LP decoding over the AWGNC, another suggestion for solving the difficult optimization problem faster was formulated in \cite{08CS} by Chertkov and Stepanov. This pseudo-codeword search (PCS) algorithm , originally stated for the continuous channel model, is based on the aforementioned relation between instantons and respective pseudo-codewords. Specifically,  if a pseudo-codeword, ${\bm p}$ corresponding to an instanton, is known, then reconstructing the respective instanton $\tilde {\bm p}$ is equivalent to maximizing the probability of the noise under the condition that the probabilities of the noise configuration counted from the zero-codeword and from the pseudo-codeword, ${\bm p}$, are identical, i.e.
\begin{equation}
\tilde{\bm p}=\left.\mbox{argmax}_{\bm n} P({\bm n})\right|_{P({\bm n})=P({\bm n}+{\bm p});\quad {\bm p}\neq {\bm 0}}.
\label{max_prob}
\end{equation}

The idea of the method of \cite{08CS} consists of throwing a sufficiently strong configuration of the noise (so that the resulting decoding is erroneous), decode it into a pseudo-codeword, and then assume that the pseudo-codeword shares an error-surface with the zero-codeword. Then the projected instanton is reconstructed using Eq.~(\ref{max_prob}), even though the noise configuration, especially after the first iteration, is not an actual instanton. This procedure is repeated until the input and the output for an iteration give the same result. It was empirically shown in \cite{08CS}  that such scheme formulated for the LP decoder outputs the sequence of noise configurations with probabilities monotonically increasing with the number of iterations and converging in a small number of iterations to an instanton. 

\subsection{Instanton search for LP decoding over the BSC}
The PCS was extended to the case of LP decoding over the BSC by Chilappagari \textit{et al.} in \cite{08CCV}. The algorithm proposed in \cite{08CCV} termed as the instanton search algorithm (ISA) is provably efficient and outputs an instanton in bounded number of steps. We summarize the algorithm below.
 
\noindent \underline{\textbf {ISA for LP Decoding over the BSC}}\\
\underline{\textit{Initialization (l=0) step}}: Initialize to a binary input vector $\mathbf{r}$ containing sufficient number of flips so that the LP decoder decodes it into a pseudo-codeword different from the  all-zero-codeword. Apply the LP decoder to $\mathbf{r}$ and denote the pseudo-codeword output of LP by $\mathbf{p}^{1}$.\\
\underline{\textit{$l\geq 1$ step}}: Take the pseudo-codeword $\mathbf{p}^l$
(output of the $(l-1)$ step) and calculate its median $M(\mathbf{p}^l)$. Apply
the LP decoder to $M(\mathbf{p}^l)$ and denote the output by
$\mathbf{p}_{M_l}$.  Only two cases arise:
\begin{itemize}
\item $w_{BSC}(\mathbf{p}_{M_l}) < w_{BSC}(\mathbf{p}^l)$. Then $\mathbf{p}^{l+1}=\mathbf{p}_{M_l}$ becomes the $l$-th step output/$(l+1)$ step input.

\item $w_{BSC}(\mathbf{p}_{M_l})=w_{BSC}(\mathbf{p}^l)$. Let the support of $M(\mathbf{p}^l)$ be $S=\{i_1,\ldots,i_{k_l}\}$. Let $S_{i_t}=S \backslash \{i_t\}$ for some $i_t \in S$. Let $\mathbf{r}_{i_t}$ be a binary vector with support $S_{i_t}$. Apply the LP decoder to all $\mathbf{r}_{i_t}$ and denote the $i_t$-output by $\mathbf{p}_{i_t}$. If $\mathbf{p}_{i_t}=\mathbf{0}, \forall i_t$, then $M(\mathbf{p}^l)$ is the desired instanton and the algorithm halts. Else, $\mathbf{p}_{i_t} \neq \mathbf{0}$ becomes the $l$-th step output/$(l+1)$ step input.
\end{itemize}

The interested reader is referred to \cite{08CCV} for a discussion of various issues that arise in the implementation of the ISA.
\section{Numerical Results}\label{section5}

This section summarizes statistics of instantons found for the $[155,64,20]$ Tanner code \cite{01TSF} performing over the BSC and the AWGNC and decoded by iterative and LP decoders. The Tanner code is a $(3,5)$ regular code whose Tanner graph has girth $8$ \cite{01TSF}.

\subsection{Instanton Statistics for the Tanner Code}
\begin{figure*}
\centering
\subfigure[]
{
    \label{InstantonBSCGalA}

\includegraphics[width=1.5in]{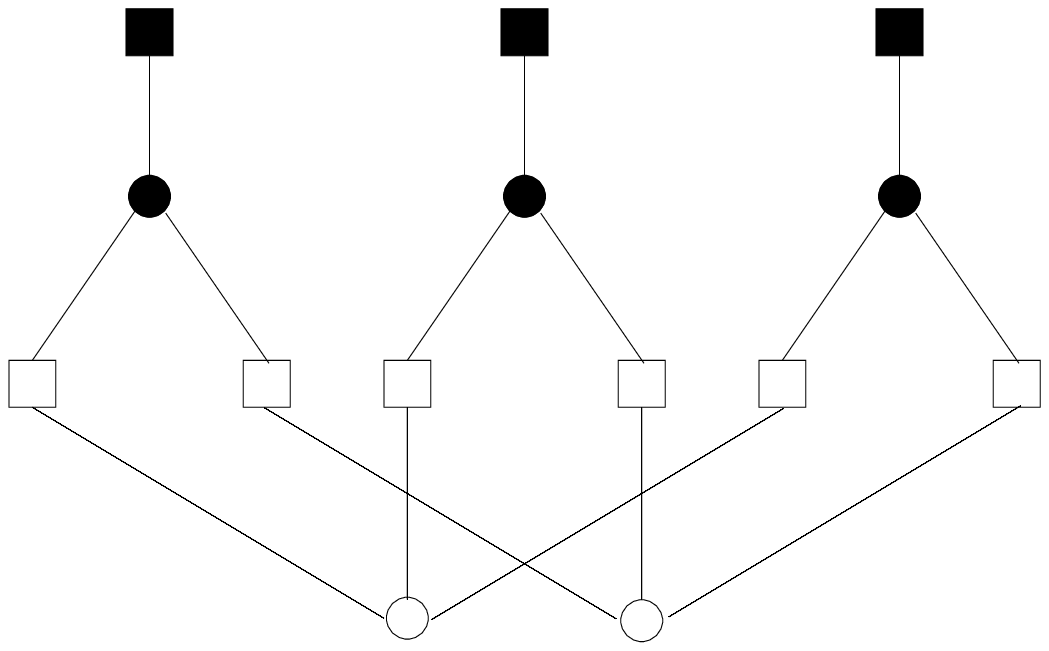}
}
\subfigure[] 
{
    \label{InstantonAWGNIterative}

\includegraphics[width=1.5in]{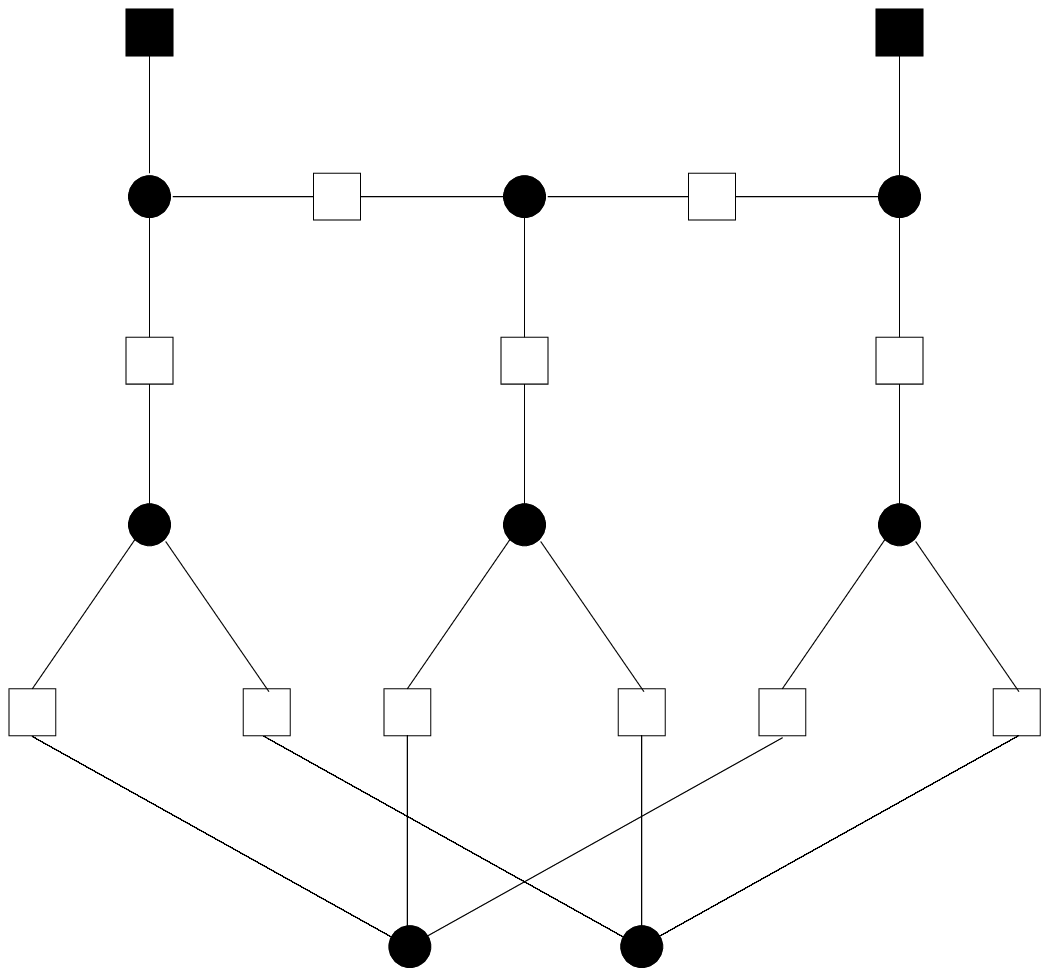}
}
\subfigure[] 
{
    \label{InstantonBSCLP}

\includegraphics[width=1.5in]{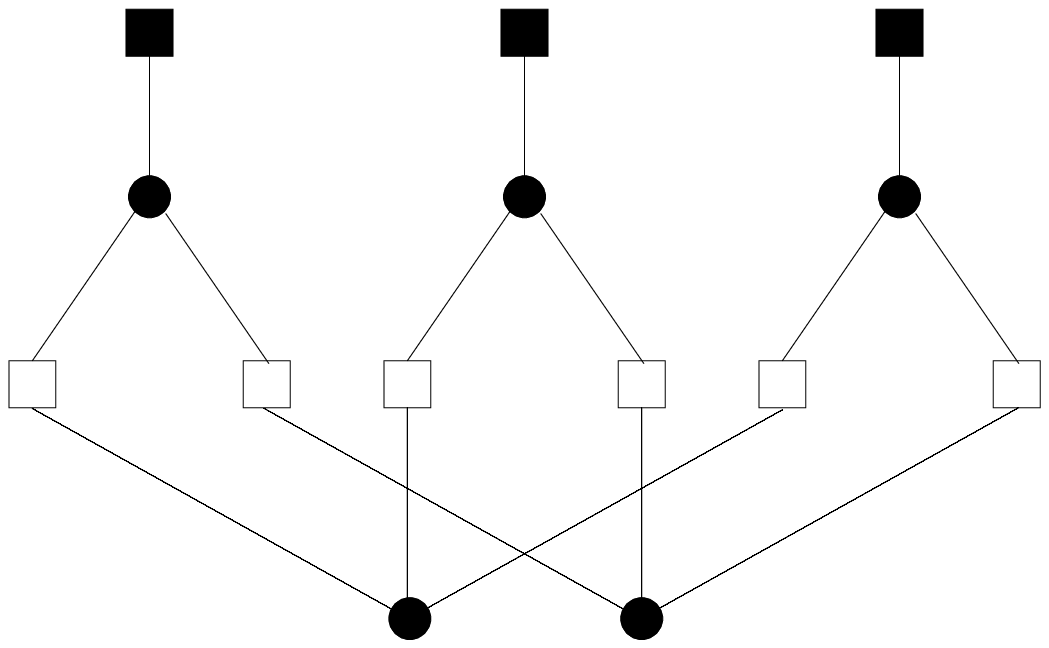}
}
\subfigure[] 
{
    \label{InstantonAWGNLP}

\includegraphics[width=1.5in]{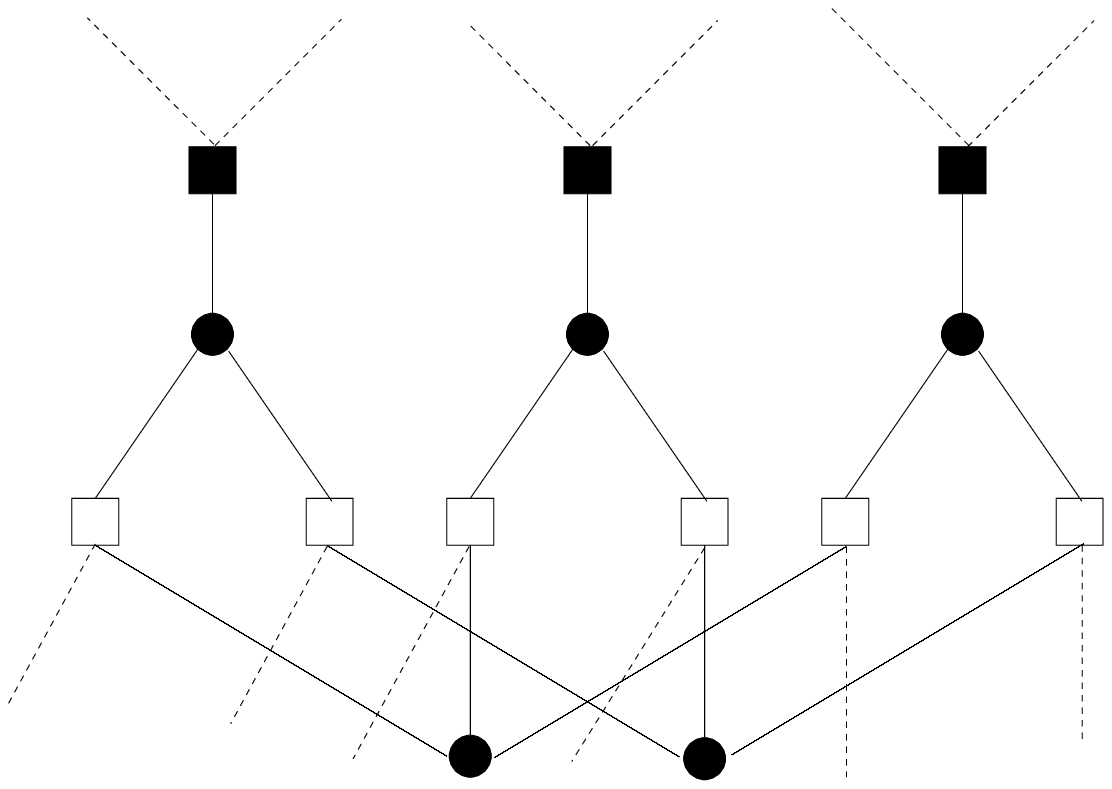}
}
\caption{Illustration of the topological structure of instanton for different channels and decoders. Note that $\square$ represents an even degree check node and $\blacksquare$ represents an odd degree check node.  \subref{InstantonBSCGalA} A (5,3) trapping set with critical number 3 for Gallager A algorithm. Here $\circ$ denotes a correct variable and $\bullet$ denotes a corrupt variable node. \subref{InstantonAWGNIterative} An (8,2) trapping set for iterative decoding over the AWGNC. \subref{InstantonBSCLP} The support of an instanton of size 5 for LP decoding over the BSC.  \subref{InstantonAWGNLP} The support of the lowest weight pseudo-codeword for LP decoding over the AWGNC. Note that the figure illustrates only the variable nodes with the largest components.} \label{illustration}
\end{figure*}

\textbf{Gallager A algorithm:}
The most dominant trapping set in the error floor domain is the $(5,3)$ trapping set which has critical number $3$. There are a total of $155~(5,3)$ trapping sets each of which has an instanton of weight $3$ \cite{06CSV} (see Fig. \ref{InstantonBSCGalA}). There are $465$ $(4,4)$ trapping sets each with critical number $4$. Hence, the slope of the FER curve in the error floor region is dominated by the $(5,3)$ trapping sets and it is equal to $3$. The trapping sets for the Gallager A algorithm are found by a combination of simulations and combinatorial considerations (see \cite{03Richardson,06CSV} for more details).

\textbf{Iterative BP:} The instantons for 4 iterations decoder were analyzed by the instanton-amoeba method in \cite{05SCCV}. The 3 lowest instantons were found, all of which contained a specific characteristic 12 bit structure. It turns out that this bit structure is what is responsible for errors even for very large number of iterations \cite{06SC}. MC simulations show that the error floor asymptotic for the Tanner code under iterative decoder with large number of iterations is determined by these structures (resulting in effective distance of 12.45 \cite{06SC}). All the trapping sets corresponding to the lowest weight instantons contain an (8,2) trapping set which is shown in Fig. \ref{InstantonAWGNIterative}. For more details on the FER curves for different number of iterations and relevant discussion, the interested reader is referred to \cite{06SC}.

\textbf{LP decoder over BSC:} The ISA described in Section \ref{section4} found $155$ distinct instantons of size $5$ (the corresponding pseudo-codewords have BSC weight $9$). The support of each of these instantons is a $(5,3)$ trapping set shown in Fig. \ref{InstantonBSCLP} (from the symmetry of the Tanner code it can be verified that there are exactly $155$ such structures present in the Tanner graph). The ISA also discovered higher weight instantons (see \cite{08CCV} for more details), but the instantons of size $5$ are the most dominant ones in the error floor region.

\textbf{LP decoder over AWGNC:} The PCS algorithm of \cite{08CS} found many low-weight pseudo-codewords ($16.4037$ being the least weight pseudo-codeword as found by the PCS). The weighted-median noise configurations (instantons) (see \cite{08CS}) corresponding to various low-weight pseudo-codewords have high noise at $5$ variable nodes corresponding to the $(5,3)$ trapping sets. In fact, the respective BSC weight $9$ pseudo-codewords have low weight on the AWGNC also (but not the absolute lowest!). The support of each of the lowest-weight pseudo-codewords is large but the components in the variable nodes corresponding to the (5,3) trapping set have maximum value (illustrated in Fig. \ref{InstantonAWGNLP}). 

An important insight gained from this comparison is that the decoding failures for various algorithms on different channels are closely related and are dependent on only a few topological structures. These relations can be exploited to find instantons for a given decoder on a given channel based on the knowledge of instantons for another already analyzed decoder,  which can even be performing over another channel. This relation is also suggestive for design of a better code, the idea substantiated in the next subsection.

\subsection{Code Design for Increasing the Smallest Instanton Size}

In \cite{07KS,08CNVM2}, it was shown that the minimum pseudo-codeword weight (for LP decoding) and the minimum critical number (for Gallager A/B decoding) of a code increase with the increase in the girth of the underlying Tanner graph. While girth optimized codes are known to perform better in general, the code length and degree distribution place a fundamental restriction on the best achievable girth. Observing that the instantons for different decoding algorithms performing over different channels have a common underlying topological structure (e.g. the $(5,3)$ trapping set in the case of the $[155,64,20]$ code), it is natural to discuss design of a similar but new code which excludes these troublesome structures.  In fact, this suggests a natural code optimization technique with an improved instanton distribution. Starting with a reasonably good code (constructed either algebraically or by the progressive edge growth (PEG) method \cite{05HEA}), we find the most damaging instantons and their underlying topological structure. We then construct a new code avoiding such subgraphs (either by swapping edges, by increasing code length, or utilizing a combination of both). We iterate this procedure till the code can no longer be optimized or reaching a computationally unbearable complexity.

For the Gallager A decoding, it has been proved in \cite{08CKV} that codes with Tanner graphs of girth $8$ which avoid the $(5,3)$ trapping set and weight $8$ codewords can correct all the error patterns of weight $3$ or less. While proving a similar result might be difficult for the iterative decoder over the AWGNC and the LP decoder, such considerations nonetheless play a role in our code design strategy. An algorithm, suggesting construction of a code meeting the Gallager A-related conditions, was provided in \cite{08CKV}. This algorithm can be seen as a generalization of the PEG  algorithm \cite{05HEA}. Given a list of forbidden subgraphs, at every step of the algorithm, an edge is established such that the resulting graph at that stage does not consist of any of the forbidden subgraphs. (The PEG algorithm is a special case forbidding cycles shorter than a given threshold.)

\begin{figure}
\centering
\includegraphics[width=3.2in]{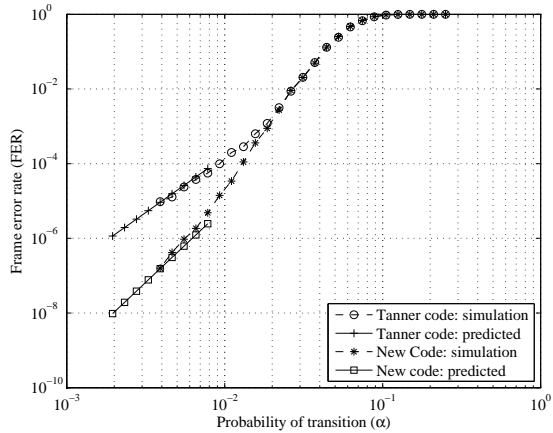}
\caption{Comparison of the FER performance of the Tanner code and the new code under the Gallager A algorithm. Plotted also is the asymptotic prediction made using the statistics of lowest weight instantons.}
\label{BSCPerformanceGalA}
\end{figure}

Using the algorithm proposed in \cite{08CKV}, we constructed a new code of length $155$ with uniform left degree $3$ and with most check nodes with degree $5$. By construction, this code avoids $(5,3)$ trapping sets. This results in a steeper  FER slope of $4$ in the error floor domain under the Gallager A decoder, as shown in Fig. \ref{BSCPerformanceGalA}. The dominant trapping set for the new code with critical number $4$ is the $(4,4)$ trapping set (an eight cycle) and has multiplicity $662$. Fig. \ref{BSCPerformanceGalA} also shows the predicted performance at very low $\epsilon$ (the method to predict the error floor performance using the trapping set statistics is described in detail in \cite{06CSV}). 

\begin{figure}
\centering
\includegraphics[width=3.4in]{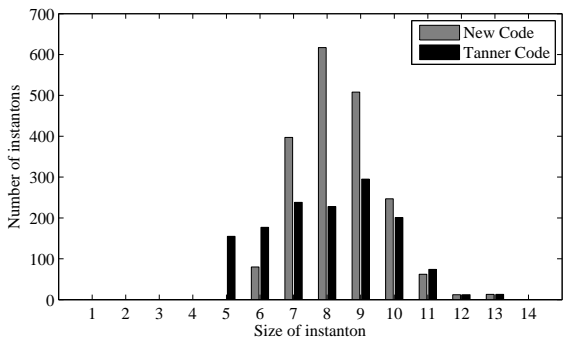}
\caption{Instanton weight distribution for the Tanner code and the new code for LP decoding over the BSC as found by running the ISA 2000 times.}
\label{OldvsNewBSC}
\end{figure}

\begin{figure}
\centering
\includegraphics[width=3.4in]{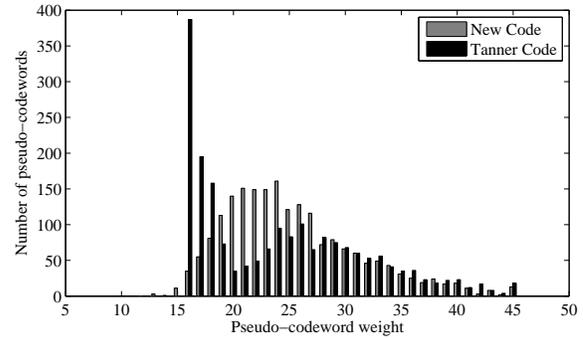}
\caption{Pseudo-codeword weight distribution for the Tanner code and the new code on the AWGNC as found by running the PCS algorithm 2000 times.}
\label{OldvsNewAWGN}
\end{figure}

\begin{table*}
\renewcommand{\arraystretch}{1}
\centering \caption{Instanton statistics obtained by running the ISA
with 20 random flips and 2000 initiations for the Tanner code and
the new code.}\vspace{10pt} \label{table:parameters2}
\vspace{-0.2in}
\begin{tabular}{|c|c|c|c|c|c|c|c|c|c|c|}
\hline
\multicolumn{2}{|c|}{\multirow{2}{*}{Code}} & \multicolumn{9}{|c|}{Number of instantons of weight}\cr \cline{3-11}
                                                                                        \multicolumn{2}{|c|}{}& 5 & 6 & 7 & 8 & 9 & 10 & 11 & 12 & 13\\ \hline
\multirow{2}{*}{Tanner code} &Total&715&194&248&230&295&201&74&10&1 \cr \cline{2-11}
                             &Unique&155&177&238&228&295&201&74&10&1 \\ \hline
\multirow{2}{*}{New code} &Total& &106&409&622&508&247&62&11&1 \cr \cline{2-11}
                          &Unique& &80&397&617&508&247&62&11&1 \\ \hline
                                            \end{tabular}

\end{table*}

\begin{figure}
\centering
\includegraphics[width=3.2in]{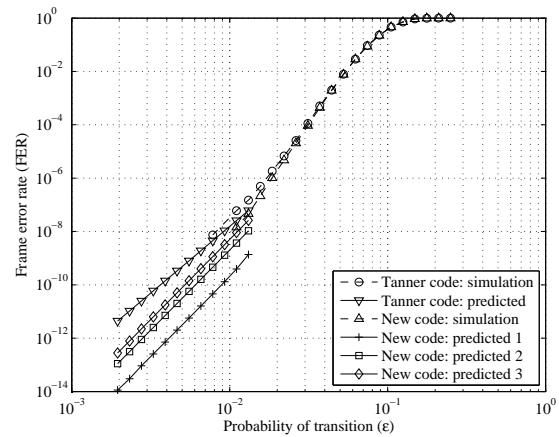}
\caption{Comparison of the FER performance of the Tanner code and the new code under LP decoding over the BSC. Plotted also is the asymptotic prediction using instanton statistics for the Tanner code. For the new code, as the total number of instantons of weight six is unknown, different curves are plotted (labeled 1,2,3) assuming different (200,2000,5000 respectively) number of instantons of weight six.}
\label{BSCPerformanceLP}
\end{figure}

The minimum weight instanton for LP decoding of the new code over the BSC found by the ISA is $6$ (we have independently verified that the code is in fact capable of correcting error patterns with up to five errors by exhaustive search). Table \ref{table:parameters2} shows the instanton statistics for the Tanner code and the new code found by running the ISA with 20 random flips and 2000 times. The statistics of number of unique instantons for the two codes as a histogram is illustrated in Fig. \ref{OldvsNewBSC}. The pseudo-codeword weight distribution for LP decoding over the AWGNC for the two codes is shown in Fig. \ref{OldvsNewAWGN}. The FER performance of the Tanner code and the new code under the  LP decoder over the BSC is shown in Fig. \ref{BSCPerformanceLP}. While all the lowest-weight instantons for the Tanner code have been found by the ISA, the same cannot be said of for the new code \footnote{The standard way to find out whether our instanton search exhausted all the unique configurations is as follows. Assume that there are $N$ unique instantons of a given weight and in each trial ISA finds all of them with equal probability. To estimate the number of ISA runs required for finding all the $N$ instantons, one notice that if $N-1$ instantons are already found the number of trails required to find to the last instanton is $\approx  N$. If all but two instantons are already found the number of ISA trials required is $N/2$. Therefore, the average number of ISA trials required to find all the instantons is estimated as $N + N/2 + N/3 + \cdots N/(N-1) + 1 = N(1 + 1/2 + 1/3 + \cdots + 1/N)$ turning to $N\ln N$ at $N\to\infty$, i.e. $N\ln N$ trials  ISA reliably finds $N$ instantons.}. Hence, we can only predict the slope of the FER curve in the error floor region and not the exact value. This can be remedied by running more trials of the ISA or by studying the automorphism group of the code and exploiting the structure of the code to find the multiplicity of the lowest-weight instantons. In Fig. \ref{BSCPerformanceLP}, we have plotted the predicted FER curve assuming different values (200, 2000 and 5000 respectively) for the number of instantons of weight six.  All the above statistics illustrate the superiority of the new code.  

\section{Conclusion}\label{section6}

In this paper, we presented a comprehensive description of various instanton based techniques for the analysis and reduction of error floors of LDPC codes. The most powerful method discussed is the pseudo-codeword/instanton search algorithm, designed specifically for the LP decoder. Using the instanton-based technique for analysis of sample (intermediate size) codes, e.g. $[155,64,20]$ Tanner code, we conclude that the underlying topological structures of the most probable instanton, found for the same code but different channels and decoders, are related to each other. Understanding of the graphical structure of the instanton and its relation to the decoding failures leads to a method to construct codes whose Tanner graphs are free of these structures. The instanton technique, applied to this code and also complemented by the direct Monte Carlo simulations, confirm the success of the new code improvement strategy. 

Future work includes: (1) refining the above techniques and applying them to longer codes, (2) developing improved semi-analytical methods for FER estimation, more specifically combining instantons and MC in order to obtain a good approximation of the entire FER curve, (3) optimization of decoders to reduce error floors, and (4) finding other combinatorial strategies for designing low error floor codes, such as judicious removal of the lines in a finite geometry leading to point-line incidence matrix free of trapping sets.
\section*{Acknowledgment}
Part of the work by S. K. Chilappagari was performed when he was a summer GRA at LANL. The work at LANL, by S.~K.~Chilappagari and M.~Chertkov, was carried out under the auspices of the National Nuclear Security Administration of the U.S. Department of Energy at Los Alamos National Laboratory under Contract No. DE-AC52-06NA25396. B.~Vasic and S.~K.~Chilappagari would like to acknowledge the financial support of the NSF (Grants CCF-0634969 and IHCS-0725405) and Seagate Technology. M.~G.~Stepanov would like to acknowledge the support of NSF grant DMS-0807592. The authors would like to thank A.~R.~Krishnan for providing the modified code and the anonymous reviewers for their suggestions.


\end{document}